# Some notes concerning the prediction of the amplitude of the two solar activity cycles

A. G. Tlatov

*Kislovodsk Solar Station of the Pulkovo Observatoy,, Russia*
*E-mail: solar@narzan.com*

The parameter G, which is determined from the general number of sunspots groups **Ng** according to the daily observations **G=(Σ1/Ng)$^2$,** is offered. This parameter is calculated for the days when there is at least one sunspots group. It characterizes the minimum epoch solar activity. Parameter G mounts to the maximum during the epoch close to the minimal activity of sunspots.

According too the data of the sequence of sunspots group in Greenwich-USAF/NOAA observatory format, observation data of Kislovodsk solar station and also daily Wolf number the changes of parameter G during 100 years were reconstructed. It is demonstrated in the paper that parameter G's amplitude in minimal solar activity *n* is linked with the sunspot cycle's amplitude $W_{n+1}$. The 24$^{th}$ activity cycle prediction is calculated, which makes $W_{24}$=135 (±12).

## Introduction

The solar activity prognosis is a topical issue, relating to the applied aspects of solar-terrestrial relationship and the fundamental nature of the solar magnetic cyclicity. Among the different methods of the amplitude of the sunspots activity cycle prognosis, the most successful are those where the solar activity precursors are used [Li et al., 2001]. The results of geomagnetic activity's observation [Ohl, 1970], large-scale magnetic fields [Tlatov & Makarov, 2005], polar magnetic field [Svalgaard, et al., 2005] and others are used as a basis for such methods.

Along with that the sunspots are the most long-term observations of the solar activity. Nowadays there have been discovered the methods that allow to appraise the amplitude of the sunspots activity cycle in accordance with the sunspots data. The most famous of them are the Gnevishev- Ohl' rule, the amplitude-period method, maximum-minimum method and others [Hathaway et al., 1999]. As a rule, these methods allow to appraise the amplitude of the following sunspots activity cycle relying upon the characteristics of the current cycle or minimum epoch.

In this paper the method of appraising the amplitude of the sunspots activity cycle is considered, according to the number of the sunspots groups and daily Wolf number.

## The Analysis of the Sunspots Groups' Observation Data

The numbers of the sunspots groups in Royal Greenwich Observatory - USAF/NOAA sunspot data were used as a basis of the analysis. These data were taken from the web-site http://solarscience.msfc.nasa.gov/greenwch.shtml along with the data of sunspots of Kislovodsk solar Station. Let's introduce parameter **G**, which is determined from the general number of sunspots groups **Ng** according to the daily observations $G=(\Sigma 1/N_g)^2$. This parameter is calculated for the days when there is at least one group of sunspots in sight. So, for the days when the number of groups is equal to 0, 1, 2, 3 … , the parameter G meanings amount to 0, 1, 1/4, 1/9 … correspondingly. Figure 1 depicts the parameter G behaviour resulting from the monthly average meanings and smoothed by means of the sliding average method by 12 points.

The **G** parameter has its maximum in the minimum activity epoch. The amplitude of the **$G_{n-1}$** parameter precedes the amplitude of the sunspots activity cycle **n**.

Figure 2 shows regression dependence between the amplitude of the **$G_{n-1}$** parameter and the amplitude of the sunspots activity in cycles **n** **$W_n$**. For cycles 16-23 this correlation may be presented as: **$W_n=-30(\pm 21)+570(\pm 71)G_{n-1}$**, the standard deviation $\sigma=12,2$ and coefficient of correlation R=0,956. The 24$^{th}$ activity cycle prognosis according to index G amounted to $W_{24}=135(\pm 12)$.

Closely related result was received at Kislovodsk solar station. The observations of the Kislovodsk solar station have been carried on since 1954. On Figure 3 changes of G parameter throughout the period of 1954-2006 are presented. In spite of the different level of meanings conditioned by different systems of the sunspots groups' calculations, one observes a considerable accordance between the numbers of Kislovodsk and Greenwich observatory format data. Before the 19$^{th}$ activity cycle one observes a slight peak of G parameter, and then it is followed by two close amplitude cycles.

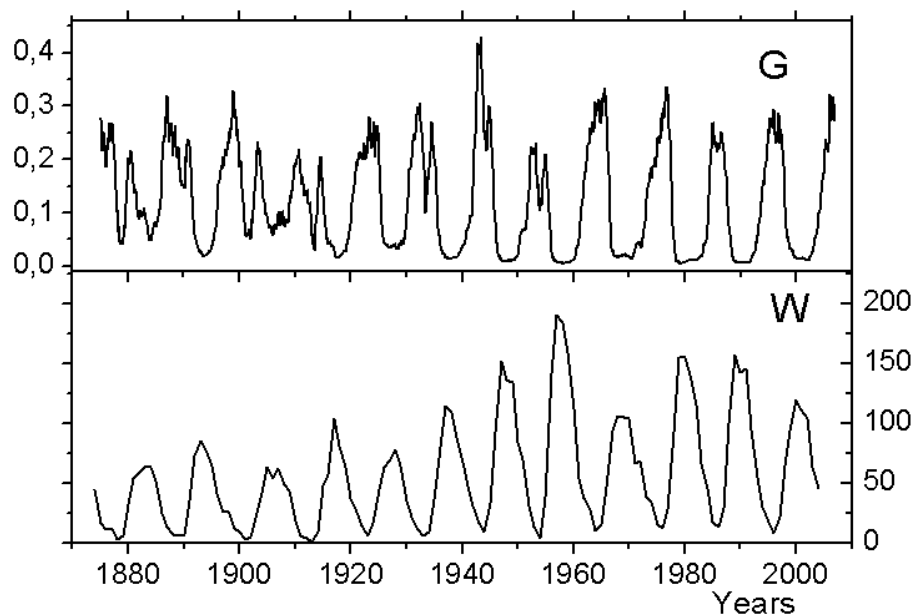

**Figure 1.** *(above)* the parameter $G = \sum 1/N_g^2$ behaviour resulting from the monthly average meanings and smoothed by means of the sliding average method by 12 months relating to the data of Greenwich observatory format.
*(below)* average annual Wolf numbers.

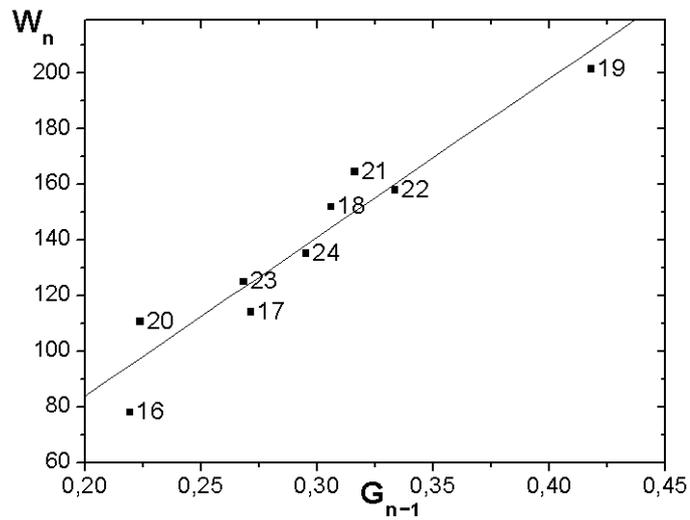

**Figure 2.** The amplitude of the sunspots activity cycles $W_n$ as a function from the $G_{n-1}$ parameter.

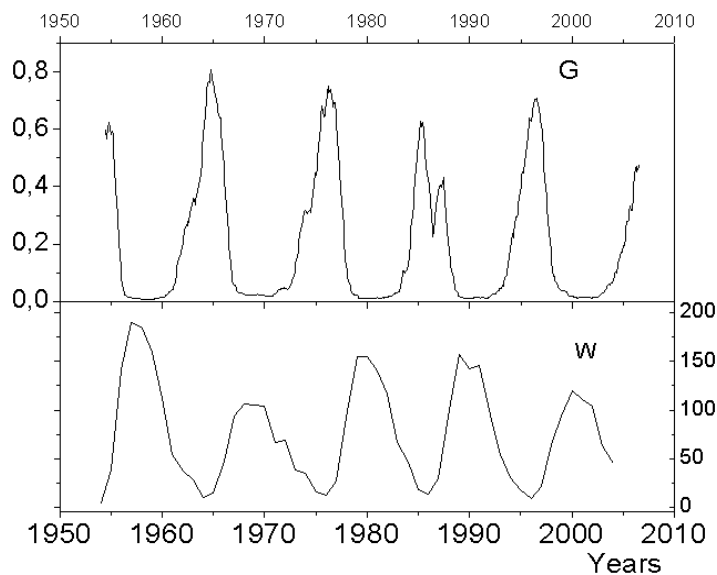

**Figure 3.** *(above)* the variations of the parameter G according to the data of the groups of sunspots of Kislovodsk solar station. *(below)* average annual Wolf numbers.

**Analysis of the sunspots' data index Rz**

In order to describe solar activity, an index of the sunspots **Rz** is widely used. The index itself was introduced by Rudolf Wolf.

The relative sunspots index is linked with the numbers of number of the sunspots groups Ng by the correlation $Rz = k \cdot (10 \cdot Ng + n)$, where k- correctional factor for an observation, n- the number of the sunspots in groups. Let's consider the following procedure of getting an index analogous to G with respect to daily index Rz. For Rz meanings transcending 7, we will introduce a filter function equal to : $g = 50*(1.0 - \exp(-z + 1.0 - \exp(-z))) + 7$, where $z = (Rz - 18)/20$; and we'll put $g = 50$ for $Rz < 8$. Further we'll use the procedure analogous to building of the G index. Thus, we'll get the monthly average sums, divide them on the amount of days per month and raise to the fourth power. So, this procedure defines the GW $24^{th}$ activity cycle prognosis parameter which results from the daily Wolf numbers. It should be noted that before 1976 period a simpler transformation can be used. In order to do it, one should take an integer part of the W/8 meaning. Figure 4 reflects the behaviour of the received GW parameter, smoothed by 18 months and Wolf numbers. Figure 5 shows the regression between the amplitude of the sunspots cycles and parameter GW. The dependence can be represented in the formula: $W_n=-14.6(\pm 27)+5.4(\pm 1)GW_{n-1}$, the correlation level is R=0,865 and a standard deviation σ=22. . The amounted to $W_{24}=140(\pm 22)$, that is close to the index G meaning.

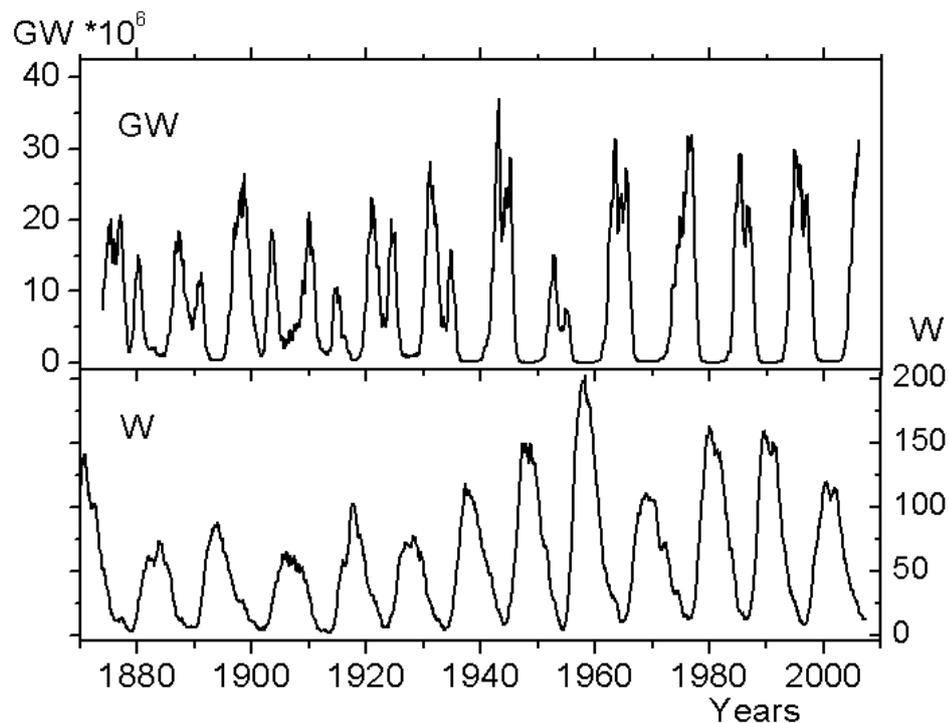

**Figure 4.** *(the upper panel)* GW index, received from the daily meanings of Wolf numbers, smoothed by 18 months. *(lower panel)* smoothed Wolf numbers.

**Discussion**

The method offered above allows to build solar activity indices, which reach their maximum in the minimum epoch, and the amplitude of which precedes the amplitude of the sunspots cycle. In contrast to the indices of large-scale magnetic field, which occur 5-6 years before the solar activity (Tlatov & Makarov, 2005), the amplitude of the new indices G and GW outdistances the amplitude of the sunspots cycle one and a half 11-year cycles.

The nature of such a connection is not quite clear. Probably, it id defined by the characteristic time of the sunspots magnetic field transformation, necessary for solar cyclic recurrence generation.

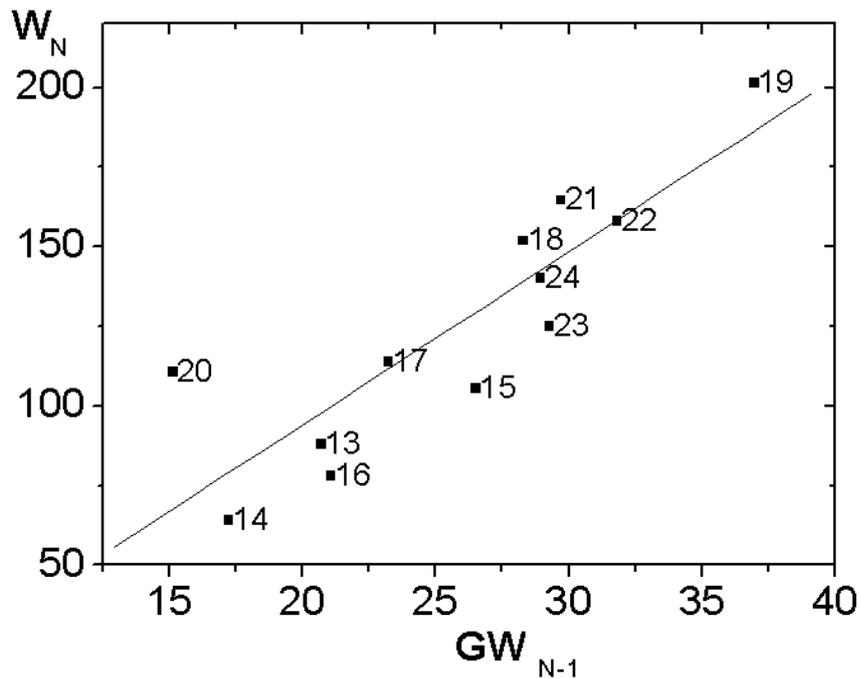

**Figure 5.** The dependence of the amplitude of activity cycles on the GW parameter. The 24$^{th}$ activity cycle prediction is calculated.

This paper was supported by the Russian Fund of Basic Researches, projects 06-02-16333